\documentclass[prl,twocolumn%
,showpacs
,preprintnumbers,amsmath,amssymb]{revtex4}

\usepackage{placeins} 
\usepackage{times}

\usepackage{here}
\usepackage{graphicx}
\usepackage{dcolumn}
\usepackage{bm}
\usepackage{paralist}

\usepackage{amsfonts}

\begin{document}


\title{Corticothalamic projections control synchronization in locally coupled bistable thalamic oscillators}
\author{J\"org Mayer$^{1}$, 
Heinz Georg Schuster$^{1}$, Jens Christian Claussen$^{1}$,  
Matthias M\"olle$^{2}$ }
\affiliation{$^{1}$Institute for Theoretical Physics and Astrophysics, 
University of Kiel, 24098 Kiel, Germany \\
$^{2}$Department of Neuroendocrinology, University of L\"ubeck, 23538 L\"ubeck, Germany}

\date{February 8, 2007}

\begin{abstract}
Thalamic circuits are able to generate state-dependent oscillations of different frequencies 
and degrees of synchronization. 
However, only little is known how synchronous oscillations, like spindle oscillations in the thalamus, 
are organized in the intact brain. 
Experimental findings suggest that the simultaneous occurrence of spindle oscillations over widespread 
territories of the thalamus is due to the corticothalamic projections, 
as the synchrony is lost in the decorticated thalamus. 
Here we study the influence of corticothalamic projections on the synchrony in a thalamic network, 
and uncover the underlying control mechanism, leading to a control method which is applicable in wide range
of stochastic driven excitable units.
\end{abstract}


\maketitle

Coupled oscillators are abundant in physics \cite{wies1,ros}, chemistry \cite{kis} and biology \cite{buc,wal,baz}. Whenever large numbers of coupled oscillators are considered, the collective behavior of the ensemble 
is of great interest \cite{ros,kur}. A widespread phenomenon in populations of periodic, noisy and chaotic oscillators 
(or maps) 
is the appearence of synchronous collective oscillations, studied theoretically \cite{kur} as well as experimentally \cite{kis,ned}. 
In neural systems the phenomenon of spike burst activity is widespread and 
of great importance \cite{muk,mol}. 
This activity is characterized by a recurrent transition between a resting state and a firing state with a burst of multiple spikes. 
Bursting is a multiple time scale effect and 
appears due to a slow process which modulates a fast subsystem \cite{muk,hind}. 

Neural 
 cells in the thalamus are known to exhibit spike burst activity during periods of drowsiness, inattentiveness and sleep \cite{des1}. Spindle oscillations, a hallmark of early sleep stages, observed in the electroencephalograhic recordings of sleeping humans as oscillations in a 12-15 Hz frequency range, are considered to be the result of synchronized spike burst activity of millions of neurons in the thalamus \cite{mol,ster}.
Despite experimental findings in thalamic slices, where the spindle oscillations propagate like a traveling wave through the network, in the intact brain spindle oscillations occur almost synchronously over widespread territories of the thalamus \cite{ster,des}. These contrasting results between in vitro and in vivo experiments occur due to the absence of corticothalamic projections (synaptic connections from the cortex to the thalamus) in a thalamic slice as, after ablation of the cortex, synchrony is lost in the thalamus \cite{con,ster,des}.

In a coupled system of oscillators,
oscillations become more and more coherent when the coupling strength is increased \cite{kur,ros}.
Here we explore
 the possibility to induce the transition from the decoherent state to the coherent state
--~in which bursts occur synchronously~-- by external signals.  For this reason, we investigate the mechanism by which corticothalamic projections control coherence of thalamic spindle oscillations. The knowledge of control mechanisms for synchrony in neural systems may help to understand the origins and mechanisms of sleep and other processes in the mammalian brain, 
and --~in the long run~-- to control them.

The paper is organized as follows: We first describe our computational model, which is a
simplified
 phenomenological
model of the thalamocortical oscillator described detailed in \cite{des1,may}. Then, in a first startup, we study burst-synchronization analytically for the case of two coupled neurons. 
Finally, we study the collective behavior of thalamic oscillators numerically, and verify the experimental observation that corticothalamic projections control synchrony in the thalamus by using a hybrid network, consisting of a computational model of the thalamus and real human slow wave EEG data as the control signal.

\begin{figure}[htbp]
\includegraphics[angle=0,totalheight=4.0cm]{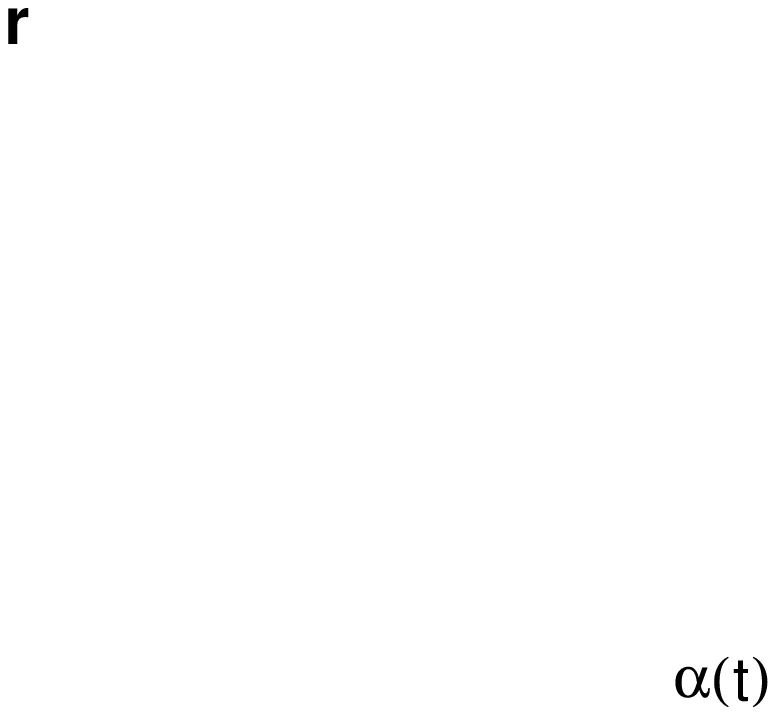}
\caption{\label{fig1} Left: Depending on $\alpha$, the system possesses either one stable fixed point or it shows bistability with a stable fixed point and a stable limit-cycle.  
Right: Bifurcation diagram of eq.\
  (\ref{lange})
 with $\alpha(t)$ considered as a bifurcation parameter.
Solid lines indicate a stable fixed point, open circles an unstable limit cycle and filled circles a stable limit cycle. 
Depending on $\alpha$, the system possesses either one stable fixed point, or it shows bistability with a stable fixed point and a stable limit cycle. The limit-cycle dissapears by an inverse sniper (saddle node into periodic orbit) bifurcation.}
\end{figure}

The dynamics of the system described in \cite{may} is mainly determined by a slow change between a bistable state where a stable fixed point and a stable limit cycle coexist, and a monostable state where only a stable fixed point exists. This behavior can be described by a slowly modulated over-damped movement of a particle in a rotating symmetric time-dependent
 double well
 potential 
(Fig.\ \ref{fig1})
\begin{equation}\label{pot}
U(r,t,\varphi)=\frac{1}{2}r^2 -  \frac{1}{4} \alpha(t)r^4 + \frac{1}{6}r^6,
\end{equation}
where $r=\sqrt{x^2+y^2}$, 
and $\varphi(t)=\omega t$ captures the oscillation
in phase (mod $2\pi$),
see 
 Ref.\ \cite{tos} for a related model.
Depending on the slow variable $\alpha(t)$, the system possess either one stable fixed point, or it shows bistability with a stable fixed point and a stable limit-cycle. 
In this paper, we restrict ourselves to very slow changes in $\alpha(t)$. In that case, the relaxation time within the wells is short compared to $\alpha(t)$, so the quasi-adiabatic approximation is applicable. 
We get a Langevin equation for the movement in $r$ direction for the single oscillator:
\begin{eqnarray}\label{lange}
\dot{r}&=&-\dfrac{\partial U(r,\alpha)}{\partial r}+I^{\rm ext}+F(t)\nonumber\\
\dot{u}&=&0.01(r (1-u)-0.14 u),
\end{eqnarray}
where 
$\alpha(t)=a(1-u)$,
$F(t)$ is a common control signal.

\begin{figure}[htbp]
\includegraphics[angle=0,totalheight=5.5cm]{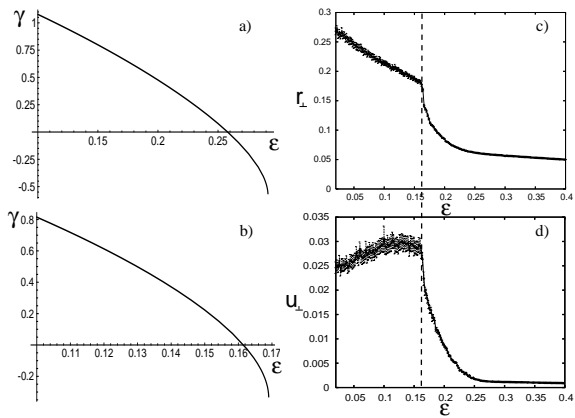}
\caption{\label{fig2}a): $\gamma$ for the case that the synchronous manifold from 
(\ref{sync}) gets disturbed, by an input spike with amplitude $0.5$ 
in the surrounding of the local maximum of $U(r,\alpha)$,
see (\ref{cond}) and Fig.\ \ref{fig1}.
b) $\gamma$ for the case that the two neurons are in different wells of the potential. 
c) and d): Variations of $r_{\perp}$ and $u_{\perp}$ as a function of $\varepsilon$. 
As the interwell jumps synchronize at $\varepsilon\simeq0.16$, 
the variations of $r_{\perp}$ (c) and $u_{\perp}$ (d) rapidly decrease at this point. 
For $\varepsilon\geq0.26$ the variations show no more significant changes as 
here the whole synchronous manifold is stable.}
\end{figure}

Following \cite{muk}, as a first step we study synchrony in two bistable elements coupled linearly via the $r$ component:
\begin{eqnarray}\label{couple1}
\dot{r}_{i}&=&-r_{i}+a(1-u_{i})r^{3}_{i}-r^{5}_{i}+I^{\rm ext}_{i}+\epsilon r_{j} \nonumber\\
\dot{u}_{i}&=&\mu (r_{i} (1-u_{i})-0.14 u_{i}),
\end{eqnarray}
where $ \epsilon > 0$ is the coupling strength, 
$a=3.1$
is chosen such that
$U(r,\alpha)$ is a double well potential for $u<0.35$, 
and $i=1,2,\;j=2,1$ respectively are the indices. $I^{\rm ext}_{i}$ is the external input, which is a Poisson distributed shotnoise with  medium rate of $1/100$ms,  refractory period of $30$ms and  pulse duration of $2$ms. 
We emphasize that the $I^{\rm ext}_{i}$ 
are stochastically independent, so $\langle I^{\rm ext}_{i},I^{\rm ext}_{j}\rangle=\delta_{i,j}$. 
The synchronized state is then given by
\begin{eqnarray}\label{sync}
\dot{r}&=&-r+a(1-u)r^{3}-r^{5}+\epsilon r\nonumber\\
\dot{u}&=&\mu (r(1-u)-bu),
\end{eqnarray}
as in this case
 $|r_{1}-r_{2}|$ and $|u_{1}-u_{2}|$ vanish for $t\to \infty$.\\
Now we consider the temporal evolution of small disturbances of the synchronous manifold. For this manner we transform to $r_{\perp}=r_{1}-r_{2}$ and $u_{\perp}=u_{1}-u_{2}$. If the system is close to the synchronous state $|r_{\perp}|,|u_{\perp}|\ll 1$, the following approximations hold: $r^{n}_{1}-r^{n}_{2}=nr^{n-1}r_{\perp}$ and $r^{n}_{1}u_{1}-r^{n}_{2}u_{2}=r^{n}u_{\perp}+unr^{n-1}r_{\perp}$. So the movement away from the synchronous state can be described by:
\begin{eqnarray}
\nonumber
\!\dot{r}_{\perp}\!\!&=&\!\!-r_{\perp}\!+3ar^{2}r_{\perp}(1-u)-ar^{3}u_{\perp}\!-5r^{4}r_{\perp}
\!+I^{\rm ext}_{\perp}\!-\epsilon r_{\perp}
\!\!\!\!\!
\label{async}
\\
\!\dot{u}_{\perp}\!\!&=&\!\!\mu ((1-u)r_{\perp}-r u_{\perp}-b u_{\perp}),
\end{eqnarray}
The stability of the synchronized state is determined by the solution of 
eqns.\ (\ref{sync},\ref{async}). 
To estimate the coupling strength at which burst synchronization occurs, we consider the quasi-adiabatic limit $\mu \to 0$ 
in eqns. (\ref{sync},\ref{async}), what represents the $r-$subsystem:
\begin{eqnarray}\label{r_(a)syn}
\dot{r}
&=&
-r+a(1-u)r^{3}-r^{5}+
\epsilon(r+\Delta r)\label{sync_r}\\
\dot{r}_{\perp}
&=&
r_{\perp}\gamma(r,u)+I^{\rm ext}_{\perp}-ru_{\perp}\label{async_r},
\end{eqnarray}
where $\gamma(r,u)=(-1+3ar^{2}(1-u)-5r^{4}-\epsilon)$, and $\Delta r$ is the deviation from the synchronous state. 
The synchronized state of (\ref{couple1}) is only stable, if the Lyapunov exponent of (\ref{async_r}) is negative. 
Further, in the quasi-adiabatic approximation we have $du/dt=du_{\perp}/dt=0$
so that $u$ and $u_{\perp}$ can be treated as parameters in eqns.\ (\ref{r_(a)syn}--\ref{async_r}).
In (\ref{sync_r}), we have to take care of the input $I^{\rm ext}$ which disturbs the synchronous manifold;
we take respect of this effect by the 
term
$\Delta r$.
In order to be able to estimate the Lyapunov exponent, we need to get $r(u)$. For this purpose we use the fact, that due to the geometry of the 
potential $U(r,t)$ in (\ref{pot}), the only area where small disturbances $r_{\perp}\ll 1$ can grow in time, is the surrounding of the local maximum of $U(r,t)$.
So we get the following conditions for $r(u)$:
\begin{equation}\label{cond}
{dU}/{dr}=0\;\textnormal{and}\;{d^{2}U}/{dr^{2}}<0
\end{equation}
As a condition for the stability of the synchronous manifold we have $\gamma(r,u)<0$. 
By setting $u=0.12$ in (13), we get $r(\epsilon)$ for the case of a maximal sensitivity of the synchronous manifold to disturbances, as in this case the system has its minimal threshold. Although the linearization is only valid for $r_{\perp}\ll 1$ we will consider the case where one neuron is in the resting state and the other neuron is in the excited state, 
which leads to values of $r_{\perp}>1$, in this case our linearization only delivers a lower bound estimation for $\epsilon$. 
 Here the approximation is supported by 
 the observation that for increasing $\epsilon$, first the jumps between the two wells synchronize and then the 
intra-well oscillations get more and more coherent.  Solving $\gamma(r,\epsilon)>0$ for $\vartriangle r=1.1$  and $\Delta r=0.5$ numerically, 
 $\gamma$ changes its sign from negative to positive at  $\epsilon\simeq 0.16$ for  $\Delta r=1.1$ and at $\epsilon\simeq 0.25$ for  $\Delta r=0.5$. If we compare Fig. \ref{fig2} we see that our estimation works also quite well for the case discussed above. \\ 
Due to this estimation of the coupling strength for the transition to the synchronized state, the variations of $r_{\perp}$ should decrease drastically at $\epsilon \to 0.16$ as here the transitions between the two wells get synchronized and should show only little changes for $\epsilon\geq 0.25$.  The numerical simulations in Fig. \ref{fig2} reproduce the results of our estimation quite well. The estimation above only gives  the values of $\epsilon$ where $r$ synchronizes, but however as $u$ obeys a linear differential equation, which gets activated by $r$, we expect $u$ to synchronize at the same value of $\epsilon$ as $r$. The numerical simulation of eqns. (\ref{couple1}) for different values of $\epsilon$ in Fig. \ref{fig2} verify this argumentation. 

\begin{figure}[htbp]
\includegraphics[angle=0,totalheight=11.cm]{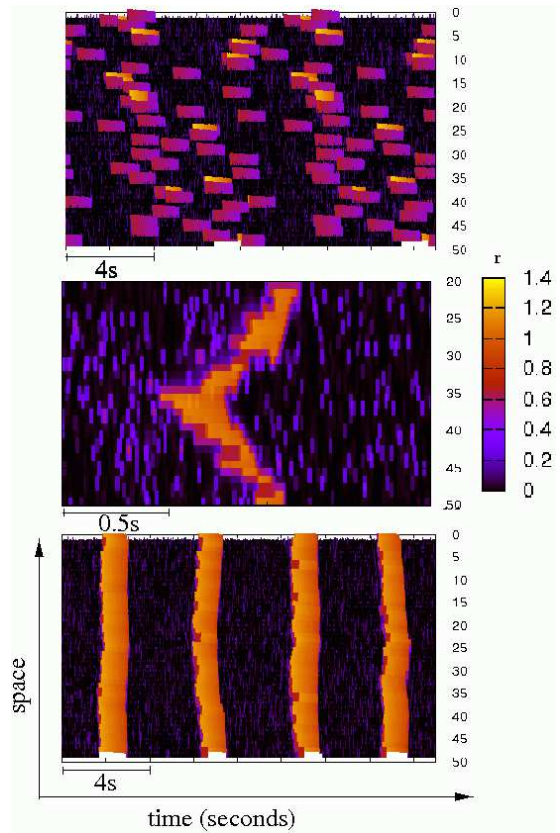}
\caption{\label{fig3} Depending on $\epsilon$ the network shows either asynchronous ($\epsilon<0.029$) or synchronous oscillations ($\epsilon>0.057$). Between these two states traveling waves occur.}
\end{figure}

Now we model the thalamic network by a 2-dimensional $50\times50$ square lattice with a nearest neighbor coupling and periodic boundary conditions.  
The network of $N$ neurons will be described by the following equation 
 \begin{equation}\label{net}
\!\left(\!\begin{array}{c} \dot{r}_{ij} \\  \dot{u}_{ij} \end{array} \!\right)  \!=\! \vec{F}(r_{ij},u_{ij})+\epsilon \left(\!\begin{array}{c} \sum_{j} G_{ij} ( r_{ij}+r_{ji} ) \\  0 \end{array} \!\right) \! +F^{\rm ext}(t),\!\!\!\!
\end{equation}
where the uncoupled dynamics of the 
$(ij)$-th
node obeys 
$\vec{F}(r_{ij},u_{ij})$ 
 given by (\ref{lange}) and the $N\times N$ matrix $G$ determines the coupling between the neurons.
For a nearest neighbor coupling its components are given by $G_{ij}=\delta_{i,j+1}+\delta_{i,j-1}$
with periodic boundaries $0\equiv N$.
$F^{\rm ext}(t)=\kappa
{\cal E}(t)$ is the common external forcing which consists 
of real human slow wave sleep EEG-data ${\cal E}(t)$.
Almost any coupling scheme can be cast into the form of
eq.\ (\ref{net}) 
by choosing the right $G$ matrix \cite{pec}, here we use a linear nearest neighbor coupling with radius 1 without self-loops.
In a first instance, we study the system without external control, such $F(t)=0$. In dependence of the coupling strength $\epsilon$ 
we observe four different kinds of collective phenomena. 
For low values of $\epsilon<0.029$ the oscillators are desynchronized.
Due to the nearest neighbor coupling for $0.029<\epsilon<0.057$ the bursts propagate like traveling waves through the network. For   $\epsilon>0.057$, burst-synchronization occurs
(see Fig.\ \ref{fig3}).
For even larger $\varepsilon$, a trivial homogeneous state (not shown) occurs.
In this work we are mainly interested in the transition of the traveling bursts to synchronous network bursts, and the possibility to induce this transition by an external control signal. As mentioned above, it was observed in thalamic slices that spindle oscillations propagate in a way similar to traveling waves in the absence of corticothalamic projections \cite{ster,des}. 
In \cite{mol} it was examined whether a comparable temporal grouping of spindle activity, coinciding with cortical slow wave oscillations can be found during human slow wave sleep. The results clearly show that also in the human sleep the spindles become synchronized by corticothalamic projections. This phenomenon was part of several numerical and experimental investigations \cite{con,bal}.

\begin{figure}[htbp]
\includegraphics[angle=0,totalheight=6.cm]{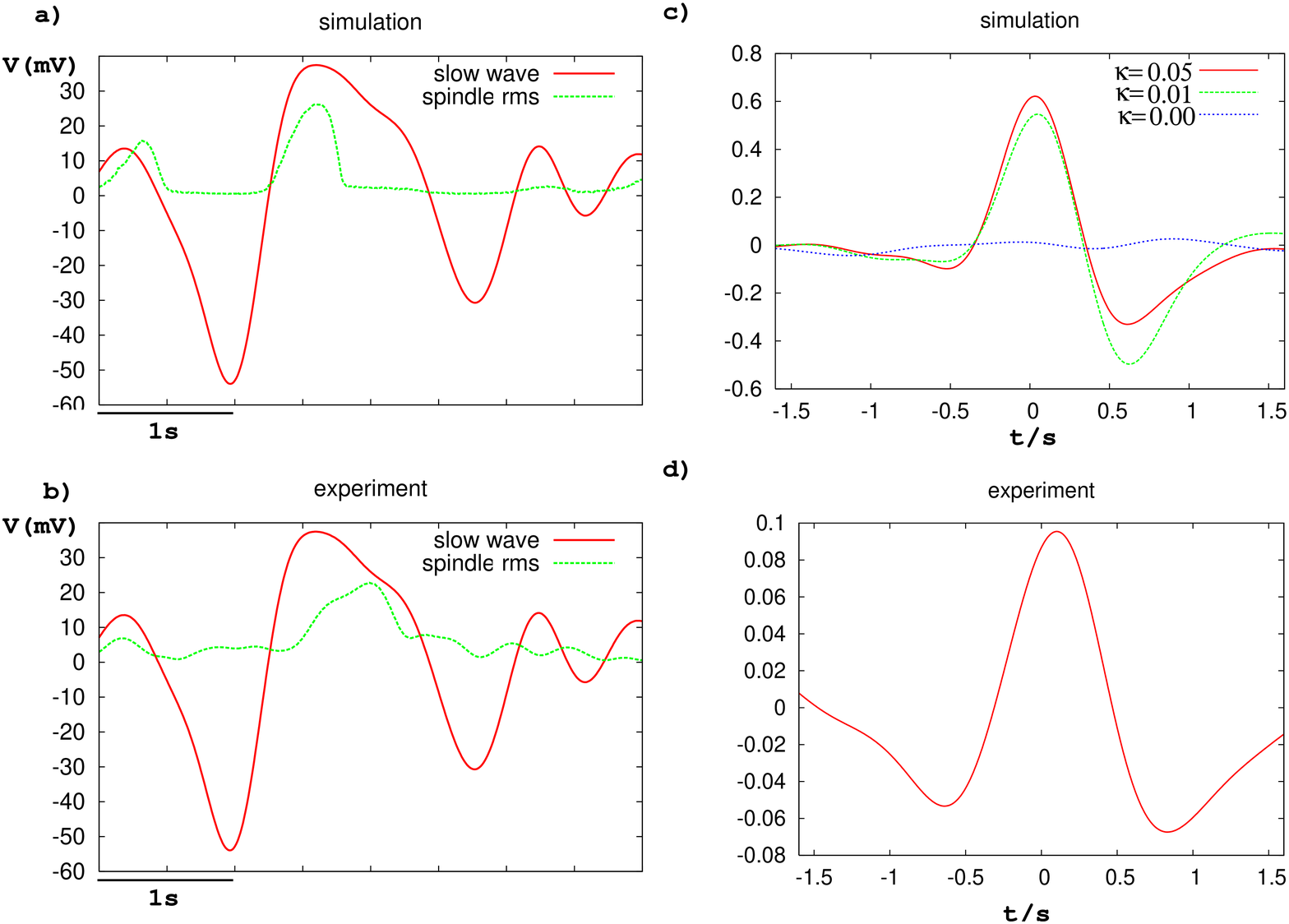}
\caption{\label{fig4} 
Correlation of meanfield network activity with driving input signal
depending on the driving parameter $\kappa$ for $\epsilon=1/25$.
a) and c): Numerical simulation of the network driven by EEG spindle sleep data as used in \cite{mol}.
For $\kappa=0.0$ correlations vanish.
b) and d): The corresponding 
experimentally obtained correlations between
spindle rms which corresponds to the of the $r_{ij}$ scaled by a factor of 30 and EEG activity.
}
\end{figure}

At this point we elucidate for a qualitative understanding of the network behaviour.
We assume the slow process $u_{ij}(t)$ to be a relaxation oscillator with relaxation time $\tau$, this relaxation process gets activated by the variable $r_{ij}(t)$, in turn $u_{ij}(t)$ inhibits excitation of $r_{ij}$ during the relaxation time $\tau$. So $u_{ij}(t)$ can be interpreted as self-inhibitory process which leads to a dead time, in which the single oscillator is not excitable. So the maximum phase difference between any arbitrary chosen oscillators is at most of the magnitude of $\tau$. 
Another way to block excitation of all $r_{ij}$ is a common sufficiently strong hyperpolarization of all oscillators, if the duration of this hyperpolarization exceeds $\tau$, all $u_{ij}(t)$ will decay back to their excitable 
ground state. That means they all have the same phase. If we further assume the input to be a Poisson distributed spike train with medium rate $\lambda$, and further assume that in the resting state $u_{ij}(t)$ gets activated by a single spike, then the mean waiting time until  $u_{ij}(t)$ gets activated is given by $1/\lambda$.
So for 
$\lambda^{-1}\ll \tau$
the $u_{ij}(t)$ can be synchronized by a sufficiently long hyperpolarization. This effect is often called phase resetting.
Of course the phase of the $u_{ij}(t)$ can diffuse by time because of the highly stochastic input. But if the phase resetting is done repetitively, then the variance of the phases of $u_{ij}$ can be confined. So we post that corticothalamic projections control coherence in thalamic slices by an open loop repetitive phase resetting. The effect of phase resetting to synchronize stochastically excited relaxation oscillators gets enforced, if the hyperpolarization is followed by a depolarization, as this increases the probability of excitation (see Fig \ref{fig4} a) and c)). 
In this sense, cortical slow waves, which consist of a hyperpolarizing half-wave followed by depolarizing half-wave, form an optimized signal to synchronize stochastic relaxation oscillators.

Finally, we verify our approach computationally in comparison with experimental data.
Consequently, we use a hybrid network consisting of a computational model of the thalamic slice, which gets real slow wave sleep EEG-data as a common input, to investigate the control of coherence in the thalamic network. 
While one might argue that the architecture of this model does not reflect nature 
in detail, as the thalamocortical projections are completely neglected, 
one reason for this approximation is the fact that the corticothalamic projections outnumber the 
thalamocortical projections by a number of magnitude \cite{ster}; 
second it was shown in cortical slices that slow waves also occur in the isolated cortex 
i.e.\ without thalamic input \cite{baz}. 
From both reasons we assume that synchronization in the thalamus mainly is controlled 
through open-loop control by the cortex. The results obtained by our hybrid network
provide a minimal model comparable with the experimental results in \cite{mol}.
Fig.\ \ref{fig4} shows that the correlation between driving EEG signal and
meanfield activity behaves similar to the corresponding experimental
result of Ref.\ \cite{mol}.

 {\sl To conclude,} we have studied the synchronizing influence of corticothalamic projections
in a thalamic network 
by a theoretical model and computationally reproduced the experiment.
As the model equations (apart from parameter choices) 
do not rely specifically on a 
neural substrate,
we anticipate 
the qualitative behavior to be generic
also for other stochastically driven 
systems composed of excitable units.
\begin{acknowledgments}
This research has been supported by the 
Deutsche Forschungsgemeinschaft
(DFG)
within SFB 654 ``Plasticity and Sleep''.
The authors thank 
Jan Born and Lisa Marshall
for intensive and fruitful discussions.
\end{acknowledgments}

\end{document}